\documentclass[amsthm,amssymb]{elsarticle}
\usepackage{graphicx}
\usepackage{epsfig}

\begin{document}
\input epsf.sty

\title{Simulation of 1+1 dimensional surface growth and lattices
gases using GPUs}

\author{Henrik Schulz}
\address{Forschungszentrum Dresden - Rossendorf, 
P.O.Box 51 01 19, 01314 Dresden, Germany}
\author{G\'eza \'Odor, Gergely \'Odor}
\address{Research Institute for Technical Physics and Materials
Science, H-1525 Budapest, P.O.Box 49, Hungary}
\author{M\'at\'e Ferenc Nagy}
\address{Research Institute for Particle and Nuclear Physics, 
H-1525 Budapest, P.O.Box 49, Hungary, and ELTE TTK, P\'azm\'any 
P\'eter 1/A, H-1117 Budapest, Hungary}

\date{\today}

\begin{abstract}
Restricted solid on solid surface growth models can be mapped onto 
binary lattice gases. We show that efficient simulation algorithms can be
realized on GPUs either by CUDA or by OpenCL programming.
We consider a deposition/evaporation model following Kardar-Parisi-Zhang 
growth in 1+1 dimensions related to the Asymmetric Simple Exclusion 
Process and show that for sizes, that fit into the shared memory 
of GPUs one can achieve the maximum parallelization speedup 
($\sim\times 100$ for a Quadro FX 5800 graphics card with respect to a 
single CPU of 2.67 GHz).
This permits us to study the effect of quenched columnar disorder, 
requiring extremely long simulation times. 
We compare the CUDA realization with an OpenCL implementation
designed for processor clusters via MPI.
A two-lane traffic model with randomized turning points is also
realized and the dynamical behavior has been investigated.
\end{abstract}

\maketitle

\section{Introduction}

Understanding basic surface growth models is an important task of statistical
physics \cite{barabasi,krug-rev}. 
The exploration of universal scaling behavior and pattern formation is a 
recent challenge as fabrication of nanostructures via self-organized 
bottom-up technology is getting widespread \cite{FacskoS}.
Simulation methods are very useful tools, since the solution of coupled 
differential equations is very difficult in general \cite{Cuerno}. 
Mapping of simple surface growth models onto binary lattice gases has 
proved to be a very useful tool to understand basic phenomena 
via nonequilibrium reaction-diffusion system \cite{odor}. 
Recently it has been shown that such mapping can be extended to higher 
dimensions \cite{asep2dcikk,asepddcikk} and more complex reactions, 
like surface diffusion ... etc. can also be treated in this way 
\cite{patscalcikk}.

The research of nonequilibrium lattice gases even in $1+1$ dimensions is of
current interest (for a review see \cite{priv97}). 
Some of them are exactly solvable, providing a framework to extend 
thermodynamical description to nonequilibrium systems
(for a review see \cite{schutz}).
Disorder, which is present in almost all real systems has been investigated 
in these systems and found to be a relevant in many cases \cite{Krug00,JSI05}.

By coupling one-dimensional lattice gases one can create the simplest
models of multi-lane traffic \cite{CSS00}, or describe the cooperative 
behavior of molecular motors \cite{CSN05}. 
These are proteins capable to carry different cargoes from one part 
of the cell to the other \cite{SW03}. Several versions of such
multi-lane system have been investigated recently, we concentrate
on the numerical simulation of a model, in which the motion of
particles in the lanes is opposite and they can jump in between the
lanes randomly \cite{J10}.
 
The computational complexity of the simulation algorithms is high enough 
to consider an implementation on modern graphics processing units 
(GPUs) since the performance of the GPUs is much higher than that of 
normal CPUs for the same cost and power supply. 
Furthermore our application has communication/computing ratio, 
that fits GPUs better than a cluster of CPUs.
To gain this performance the problem must fit to the 
architecture of GPUs, i.e. the following factors must be fulfilled:

\begin{enumerate}
\item The problem can be divided into sub-problems which can be computed 
independently (parallelization) using parallel threads.
\item The parallel threads of the algorithm do not need to communicate directly.
Data exchange can be performed using the memory (SMP approach).
\item The amount of memory needed for the computation inside each parallel 
thread is limited.
\end{enumerate}

The first of the three points is obviously essential to create a parallel 
algorithm. The second point follows from the architecture of GPUs, 
where no bus system exists between processing elements.
The third point is necessary for the performance of the algorithm only. 
Many articles in the field of statistical physics have already shown 
that there is a wide variety of simulation algorithms, which can benefit 
from GPU implementations \cite{preis,weigel,weigel2,BPP10}.

Today's GPUs are mainly produced by two manufacturers: NVIDIA and ATI. 
The architecture differs slightly \cite{nvidia,ati}, but the main 
concepts are the same: a GPU device consists of a number of multiprocessors 
(NVIDIA Fermi: 14, ATI Hemlock: 40), each equipped with several processing 
elements (NVIDIA Fermi: 32, ATI Hemlock: 16), 
an instruction block, shared memory 
and caches (see Table \ref{table1} for a summary). Each processing element 
(PE), which is also called {\it CUDA core} (NVIDIA) or {\it Stream core} 
(ATI) is a simple entity, containing arithmetic logic units, 
floating point and integer units, flow control and registers. 
The performance of PEs is limited, but due to parallelism the 
peak performance of the entire GPU is much higher than that of a 
multi-core CPU. 

For the programmer a GPU is a device, which executes {\it kernels} in 
parallel as a set of threads. There are vendor specific language extensions 
like NVIDIA's CUDA or platform independent ones like OpenCL \cite{khronos}, 
which moreover can be used to program other types of processors (DSPs, CPUs). 

The speedup, which can be achieved using one GPU (compared to running the 
simulation on a single CPU-core) is already remarkably high, but the 
run-time can still be long enough to consider multi-GPU simulations. 
On many systems it is possible to plug in more than one GPU card, but 
if more than, for example, four cards are to be used it is necessary 
to parallelize the program over a network using methods like message 
passing (MPI).

Some early results from a first GPU version have already been published 
before \cite{fzd2010}. In this article we will discuss implementations 
on different GPUs (NVIDIA and ATI) using CUDA and OpenCL for single-GPU 
versions and MPI for the multi-GPU code.
\stepcounter{footnote}
\footnotetext{64 kB can be split into 48 kB of shared memory and 16 kB of L1
cache or vice versa.}
\begin{table}
\caption{Key facts of the GPU cards from NVIDIA \cite{nvidia} and ATI
\cite{ati}.\label{table1}}
\begin{center}
\begin{tabular}{lrrr}
\hline
& NVIDIA C1060 & NVIDIA C2070 & ATI Radeon 5970 \\
& or Quadro FX 5800 & (Fermi) & (Hemlock)\\
\hline
Number of multiprocessors (mp) & 30 & 14 & 40 \\
Number of processing elements / mp & 8 & 32 & 16 \\
Clock rate & 1300 MHz & 1150 MHz & 725 MHz \\
Global memory & 4 GB & 6 GB & 2 GB\\
Shared memory per mp & 16 kB & 48 kB${}^{\thefootnote}$ & 32 kB\\
Memory clock rate & 800 MHz & 1500 MHz & 1000 MHz \\
Global memory bandwidth & 102 GB/s & 144 GB/s & 256 GB/s \\
Peak GFlop/s (single precision) & 936 & 1030 & 4640 \\
Peak GFlop/s (double precision) & 78 & 515 & 928 \\
\hline
\end{tabular}
\end{center}
\end{table}

\section{1+1 dimensional Kardar-Parisi-Zhang growth}

Surface growth in statistical mechanics has been investigated by
simple models, atomistic ones or continuum theories \cite{barabasi}. 
Understanding of several fundamental questions is still not complete.
The Kardar-Parisi-Zhang (KPZ) equation \cite{KPZeq}, motivated by
experimentally observed kinetic roughening has been the subject
of large number of theoretical studies \cite{krug-rev,HZ95}.
Later it was found to model other important physical phenomena
such as randomly stirred fluid \cite{forster77}, dissipative
transport \cite{beijeren85,janssen86}, directed polymers \cite{kardar87} 
and the magnetic flux lines in superconductors \cite{hwa92}.
It is a non-linear stochastic differential equation, which describes
the dynamics of growth processes in the thermodynamic limit specified
by the height function $h({\bf x},t)$
\begin{equation}
\label{KPZ-e}
\partial_t h({\bf x},t) = v + \sigma\nabla^2 h({\bf x},t) + 
\lambda(\nabla h({\bf x},t))^2 + \eta({\bf x},t) \ .
\end{equation}
Here $v$ and $\lambda$ are the amplitudes of the mean
and local growth velocity,
$\sigma$ is a smoothing surface tension coefficient
and $\eta$ roughens the surface by a zero-average Gaussian
noise field exhibiting the variance
\begin{equation}
\langle\eta({\bf x},t)\eta({\bf x'},t')\rangle 
= 2 D \delta^d ({\bf x-x'})(t-t') \ .
\end{equation}
Here $d$ is used for the dimension of the surface, $D$ for the
noise amplitude and $\langle\rangle$ denotes average over the noise 
distribution.

The solution of KPZ in $1+1$ dimensions is known exactly 
for different initial conditions 
\cite{kardar87,PS00,sasa05,ASQ11}, but in higher dimensions only 
approximations are available (see \cite{barabasi,odor}). 
In $d>1$ spatial dimensions due to the competition of 
roughening and smoothing terms, models described by the 
KPZ equation exhibit a roughening phase transition between 
a weak-coupling regime ($\lambda<\lambda_c$),
governed by $\lambda=0$ Edwards-Wilkinson (EW) fixed point \cite{EWc},
and a strong coupling phase.

\begin{figure}[ht]
$\begin{array}{c@{\hspace{1cm}}c}
\multicolumn{1}{l} {\mbox{\bf }} &
        \multicolumn{1}{l}{\mbox{\bf }} \\ [-0.53cm]
\epsfxsize=2.5in
\epsffile{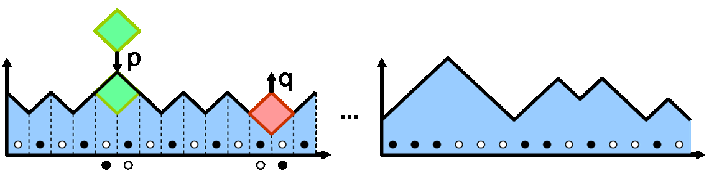} &
        \epsfxsize=1.5in
        \epsffile{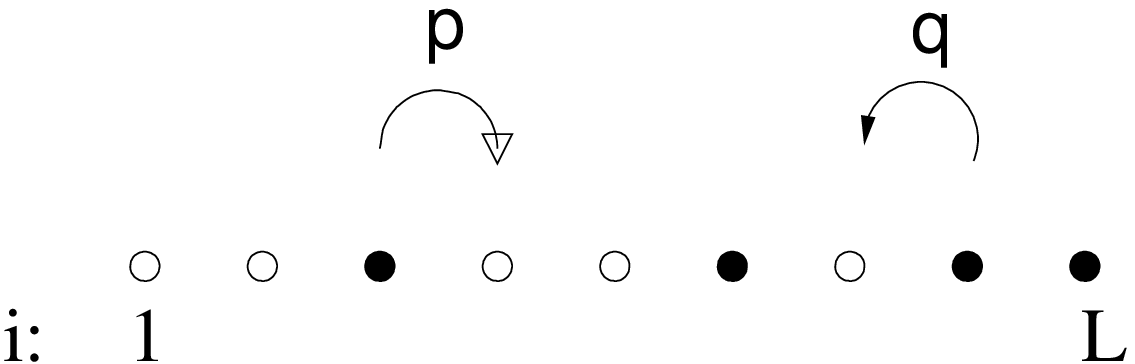} \\ [0.4cm]
\mbox{\bf (a)} & \mbox{\bf (b) }
\end{array}$
\caption{\label{kpzasep}
(a) Mapping of the $1+1$ dimensional surface growth onto the 
$1d$ ASEP model. Surface attachment (with probability $p$) and 
detachment (with probability $q$) corresponds to anisotropic diffusion
of particles (bullets) along the $1d$ base space.
(b) The ASEP model.}
\end{figure}

Mapping of surface growth onto reaction-diffusion system
allows effective numerical simulations \cite{dimerlcikk,odor}.
In one dimension a discrete, restricted solid on solid 
\footnote{The height differences $\Delta h(x,t)$ between 
neighbouring sites is finite}
realization of the KPZ growth \cite{meakin,kpz-asepmap} is
equivalent to the Asymmetric Simple Exclusion Process (ASEP) 
of particles \cite{Ligget} (see Fig.~\ref{kpzasep}(a)).
In this discrete so-called 'roof-top' or 'single-step' model the
heights are quantized and the local derivatives can take the values
$\Delta h = \pm 1$.
If we associate the up slopes with 'particles' and down slopes with
'holes' of the base lattice (see Fig.~\ref{kpzasep}(b)), the
adsorption/desorption corresponds to ASEP. 
This is a binary lattice gas \cite{KLS84,S-Z}, where particles can
hop on adjacent sites, with asymmetric rates, in case of vacancy.
The behavior of ASEP is rather well known, and variations of 
ASEP (disorder, interactions ... etc.) can be related
to variations of $1+1$ dimensional KPZ growth models.
When left/right symmetry in the hopping rates holds, we have
the so called Symmetric Simple Exclusion Process (SSEP), 
which corresponds to the EW growth model with the scaling exponents: 
$z=2$, $\alpha=\beta=0$ (i.e. logarithmic growth).

Usually the numerical solution of partial differential equations 
depends on the discretization scheme, however we used a mapping here
onto binary variables, where such discretization is not feasible. 
All gradients are approximated by '$1 / -1$' slopes and we exploited 
the spatial scale invariance of the system, which is due 
to the diverging correlation length of these models.
For CUDA we used coalesced memory access, which is optimal for
smaller system sizes that fit into the shared memory.

\subsection{Realization by CUDA}

We have programmed the one dimensional ASEP model by CUDA within the 
framework of FX 5800 (or Tesla C1060) GPU cards as follows 
\footnote{For other NVIDIA cards the algorithm is the same, one has to 
modify the hardware dependent quantities.
For example in case of Fermi card one has larger shared memory, hence
one can store larger systems.}. 
To achieve maximum performance we limited the maximum system size to
$L=16000$, hence using bit-coding technique one can store a ring 
(periodic boundary condition) within the shared memories of 
multiprocessors (in the FX5800 graphics cards there are $16$ Kbyte 
shared memory/block). 
Therefore one multiprocessor block, containing $8$ PEs, follows the 
evolution of a single chain and does not interact
with the other $30$ multiprocessors of the graphics card. 
Thus we have fast inter-PE communication during the simulations, 
and we need access of the slower card memory only at the beginning and at 
the end of a time loop.

Each multiprocessor block performs a statistically independent run,
with completely different random number sequence. This is achieved by
the intelligent striding method of {\it gpu-rng}, the pseudo-random number
generator of GNU developed for CUDA. This is in fact the parallel
version of {\it drand48}, a standard linear congruential generator 
with $48$-bit integer artihmetics \cite{gpu-rng}. This generator has
sufficiently long cycle ($\ge\sim 2^{1024}$) and the main program
pre-shifts the seeds of the individual gpu kernel parts such that they 
do not overlap with each other.

\begin{figure}[ht]
\begin{center}
\epsfxsize=70mm
\epsffile{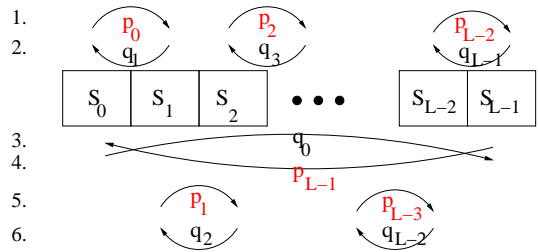}
\caption{Internal operation sequence of threads (numbered on the left) 
of a single lattice sweep of the parallel, two-sub-lattice 
update of ASEP algorithm, in case of independent jumping rates.} 
\label{parasep}
\end{center}
\end{figure}

Starting from periodic, alternating $"0/1"$ distribution, which
corresponds to a flat initial surface, we update the particle
model by the rules defined in the previous section.
The difference between the standard ASEP and the CUDA realization is
that we use parallel, two-sub-lattice updates instead of the 
random sequential one, which fits serial computers.
To allow simultaneous, multiple PE updates without particle collisions
we divide the ASEP chain (with binary state variables $\{s_i\}\in(0,1) 
\ \ , i=0,...L-1$) into even and odd blocks (see Fig.\ref{parasep}).
Since there are less processors than sites, a single sweep of the 
lattice update is done via the sequence shown on Fig.~\ref{parasep}.
First we update the even sub-lattice with forward or backward
jumps of '$1$'-s with probabilities $p_i$ or $q_{i+1}$ ($i=0,...L-2$) 
respectively. These are distributed among independent threads
of the kernel function. The best performance was obtained when the
number of threads was $256$ (for FX 5800).
Then we update the odd sub-lattice sites ($i=1,...L-3$) similarly. 
Finally we close the ring by jumping '$1$'-s between the first and 
the last sites. 
Naturally in the totally anisotropic ASEP (TASEP) case the $q_i$ 
steps are omitted, and if there is no disorder $p_i=p$ for all sites.
When all sites are updated once the time ($t$) is incremented by one 
Monte Carlo step (MCs) and we repeat these sweeps until $t=t_{max}$. 
For the stochastic cellular automaton (SCA) updates we need random 
numbers generated locally, by each exchange using {\it gpu-rng} 
in each thread.

The equivalence of ASEP with SCA dynamics and random sequential 
updates (RSU) is not evident, therefore first we compared our results
with the known results for RSU. 
At certain time steps we calculate the $h_{x}(t)$ heights from the
height differences $\Delta h = 2(s_i-1) \in(1,-1)$.
The morphology of a growing surface is usually characterized
by its width
\begin{equation}
W(L,t) = \Bigl[\frac{1}{L} \, \sum_{x=1}^L \,h^2_{x}(t)  -
\Bigl(\frac{1}{L} \, \sum_{x=1}^L \,h_{x}(t) \Bigr)^2
\Bigr]^{1/2} \ .
\end{equation}
In the absence of any characteristic length, growth processes are expected to
show power-law behavior of the correlation functions in space and height and
the surface is described by the Family-Vicsek scaling~\cite{family}
\begin{eqnarray}
\label{FV-b}
W(L,t) &\propto& t^{\beta} , \ \ {\rm for} \ \ t_0 << t << t_s \\
       &\propto& L^{\alpha} , \ \ {\rm for} \ \ t >> t_s \ . \label{FV-a}
\end{eqnarray}
Here $\alpha$ is the roughness exponent and characterizes
the deviation from a flat surface in the stationary regime
($t>>t_s$), in which the correlation length ($\xi$) has exceeded
the linear system size $L$;
and $\beta$ is the surface growth exponent, which describes the
time evolution for earlier (non-microscopic $t>>t_0$) times.
The dynamical exponent $z$, describing the space-time anisotropy,
can be expressed by the ratio
\begin{equation}\label{zlaw}
z = \alpha/\beta \ .
\end{equation}

The maximum simulation time ($t_{max}$) of runs depends on the time needed 
to achieve the steady state (saturation), hence on the system size $L$. 
We have checked this by inspecting the results on the log.-linear
scale.
Due to the parallelization among multiprocessor blocks we obtain 
$30$ independent realizations of the $W(L,t)$ data at once and there is 
an external statistical loop in the CPU code to multiply this further. 
At the end of the program the $W^2(L,t)$ data of each sample are 
averaged out by the CPU.

We have also tested the effect of the $p$ value. 
In the RSU case one can find the best KPZ scaling for $p=1$, 
but here we have to choose $p<1$, in order to introduce some 
randomness. Otherwise particles would just flip back and forth. 
A larger $p$ helps to reach the saturation earlier, but compresses 
the dynamical scaling region.
We have compared simulations with $p=0.5$ and $p=0.8$ and found 
the same scaling exponents. For the time being we fixed $p=0.8$
as a compromise between speed and dynamical scaling region size.

We have represented the state of an ASEP site ($s_i$) by the first 
bit of a character variable ($C_i$), while we used other bits of 
$C_i$ to store other, site dependent information. 
As we will see later in case of multi-lane ASEP, with quenched 
disorder this allows an effective CUDA algorithm and memory 
allocation.

In case of extremely long simulation times $t_{max} > 2^{32}$,
which is necessary to achieve the steady state of larger systems
we allowed to pass first and last elements of $s_i$ between the 
multiprocessor blocks and close the chain by the ends of the PEs
of a GPU card. In this case we simulate a single system by one
card. The communication loss, which is due to reading and writing 
of these characters to the device memory has been found to be 
small due to the fact that such events occur only once by every 
lattice sweep, i.e. with the frequency $\simeq 1/(L/30)$.

\subsection{Results for KPZ}

We have run dynamical simulations of the ASEP model with parallel SCA 
updates on rings of $L=2^9, 2^{10},...2^{14}$ sizes 
\footnote{Other lattice sizes, which are not powers of $2$ can also 
be simulated, but with a lower efficiency}. 
Each run was started from the alternating sequence of '$0$'-s and '$1$'-s, 
corresponding to the most flat surface. The probabilities of forward/backward
jumps of '$1$'-s was $p=0.8$ and $q=0$ respectively. This so-called 
Totally Asymmetric Simple Exclusion Process (TASEP) is known to exhibit
KPZ scaling in the thermodynamic limit. We sampled the evolution by
exponentially increasing time gaps: 
\begin{equation}\label{sampling}
t_{i+1} = t_i * 1.05 \ \ \ {\rm if} \ \ \ i\ne 0 \ , \ \ \ t_0=30
\end{equation}
and calculated $W^2(t_i,L)$ at these times. 
For the longest run, by $L=16000$ the saturation was reached at
$\sim 10^6$ MCs and we followed the simulations up to $10^7$ in the steady
state. This means $260$ data sampling events, according to (\ref{sampling}), 
which requires negligible time compared to the length of the 
calculations ($\sim 4$ minutes).
This ensures that the serial sampling part does not alter the 
efficiency of the parallel CUDA code.

\begin{figure}
\begin{center}
\includegraphics[height=7.5cm]{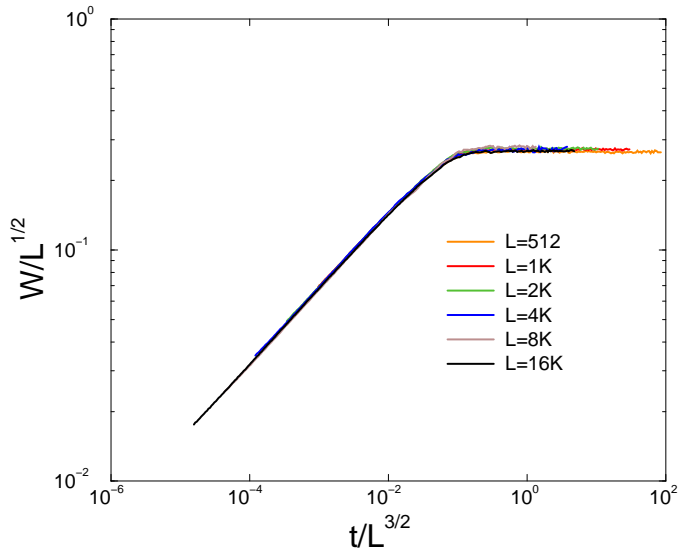}
\caption{\label{KPZtest} Scaling collapse of the $1+1$ dimensional parallel
update KPZ (ASEP) by CUDA simulation results for sizes $L=512, 1K, ...16K$.}
\end{center}
\end{figure}

By the scaling analysis we have subtracted the leading order correction
to scaling from the width data, i.e the constant value $W^2(0,L)=0.25$, 
corresponding to the intrinsic width of the initial zig-zag state.
As one can see on Fig.~\ref{KPZtest} excellent data collapse could
be achieved for different sizes with the $1+1$ dimensional KPZ 
exponents: $\alpha=1/2$ and $z=3/2$ \cite{odor}.

The parallel CUDA code run on Quadro FX 5800 card  $@ 1.3$ GHz
was compared with the same algorithm run on a CPU core 
(Intel I5 750 $@ 2.67$ GHz) and the speedup: $\times\sim 100$ was observed.
This is near to the physical limit, taking into account that the
graphics card has $240$ PE-s with twice slower clock rate.
This is the consequence of independent threads, i. e. the run-time 
data are packed in the fast shared or constant memories.
The time needed to run $30$ realizations of a $L=1024$ sized system
up to $t_{max}=10^6$ MCs on a Quadro FX 5800 card is only $4.2$ seconds.

\subsection{Results for KPZ with quenched disorder}

Having confirmed the validity of KPZ scaling for the SCA CUDA algorithm 
we introduced quenched, site-wise disorder in the hopping rates, 
with the bimodal distribution:
\begin{equation}\label{bimodal}
P(p_i) = (1-Q) \delta(p_i-p) + Q \delta(p_i - r p) \ \ ,
\end{equation}
i.e. we reduced the probability of exchange by a factor $r$
at random sites selected with probability $Q=0.5$ (here $\delta(x)$ 
denotes the Kronecker delta function). When we allow backward
jumps as well we use the same distribution for $P(q_i)$.
The sites with reduced hopping rates were tagged by the 
second/(third) bits of the $C_i$ variable corresponding to site $i$.

The TASEP particle model with quenched, site-wise disorder (Q-TASEP)
corresponds to KPZ growth with columnar disorder (QCKPZ):
\begin{equation}
\label{QCKPZ-e}
\partial_t h({\bf x},t) = v + \sigma\nabla^2 h({\bf x},t) + 
\lambda(\nabla h({\bf x},t))^2 + \eta({\bf x}) \ .
\end{equation}
i.e. the noise field $\eta(x)$ is constant in time and independent
of height.
The Q-TASEP model was investigated by scaling arguments in \cite{Krug00} 
and emphasized that "numerical confirmations for the coarsening dynamics 
in case of site-wise disorder would be most welcome".
We have done this by Monte Carlo simulations of the parallel update SCA
version of Q-TASEP. 

\begin{figure}
$\begin{array}{c@{\hspace{1cm}}c}
\multicolumn{1}{l} {\mbox{\bf }} &
        \multicolumn{1}{l}{\mbox{\bf }} \\ [-0.53cm]
\epsfxsize=2.1in
\epsffile{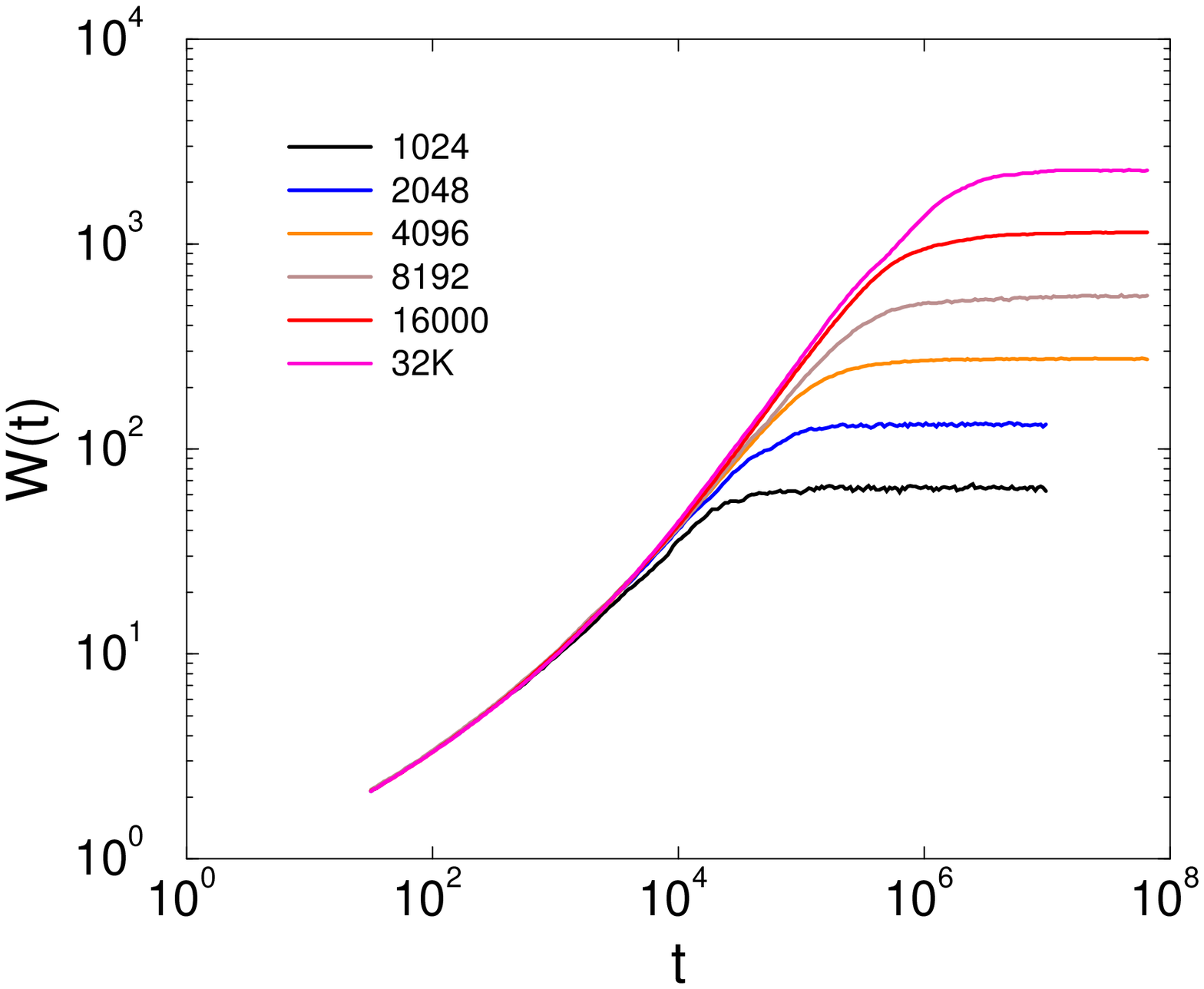} &
\epsfxsize=2.1in
\epsffile{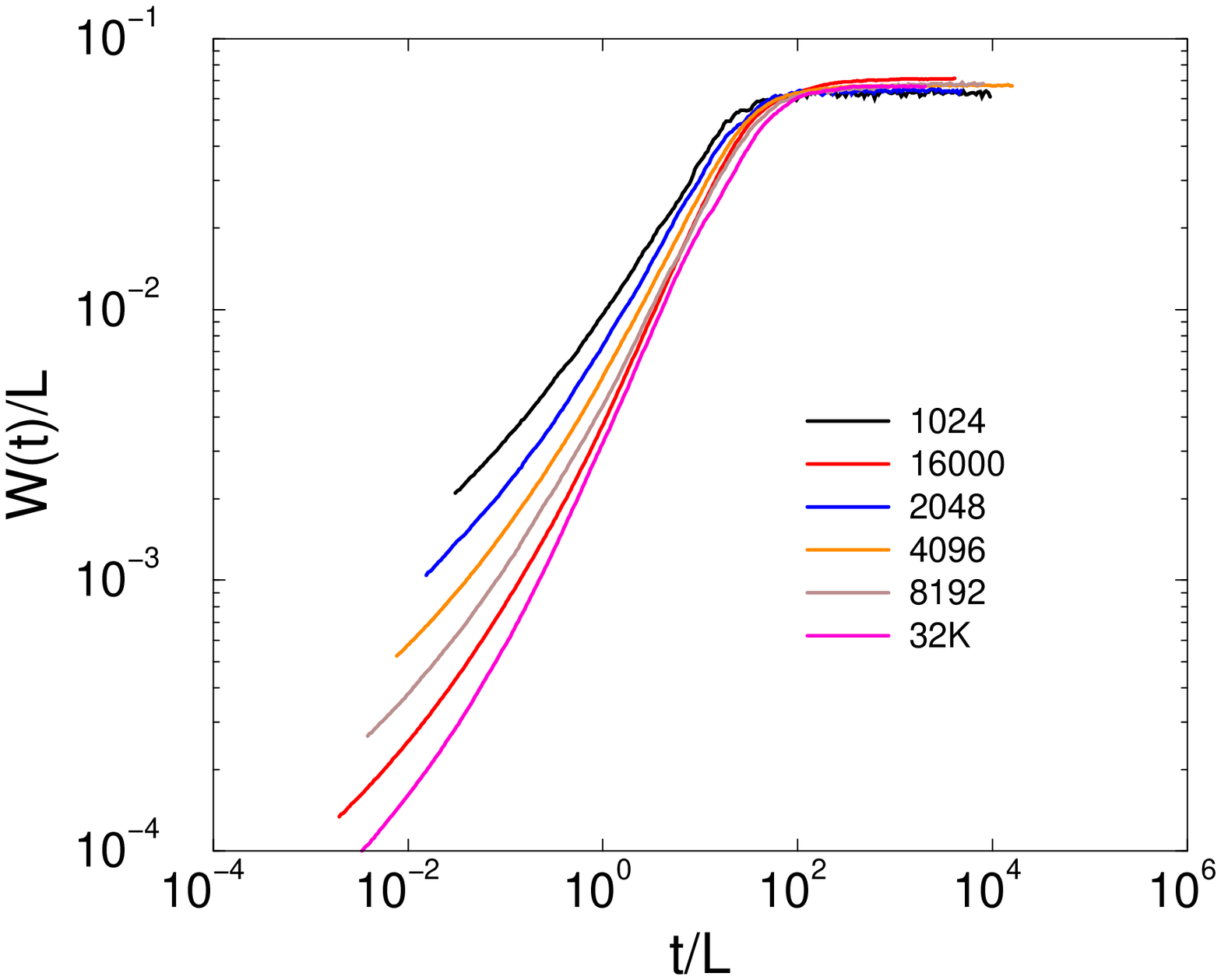} \\ [0.5cm]
\mbox{\bf (a)} & \mbox{\bf (b) }
\end{array}$
\caption{\label{q-kpz} 
(a) Surface width of QCKPZ (Q-TASEP) by CUDA simulations
for $L=1K, 2K,...,32K$ (bottom to top).
(b) The same, rescaled with the exponents $\alpha=1$, $z=1$.}
\end{figure}

As Fig.~\ref{q-kpz}(a) shows we followed the evolution of $W(t,L)$
beyond saturation for sizes: $L = 2^{10},2^{11},2^{12}, ..., 2^{15}$.
The number realizations, with independent disorder distribution
was $120$ for each size. 
The speedup compared to the serial code on the reference CPU was 
about $\times 45$, depending on the parameters slightly.
On Fig.~\ref{q-kpz}(b) we can see the same data rescaled by
the exponents $\alpha=1$, $z=1$ according to Eqs. (\ref{FV-b}) and (\ref{FV-a}). 
The data collapse is reasonable if we take granted logarithmic 
corrections to scaling
and we can confirm the results of \cite{Krug00} indeed.
More detailed analysis will be published elsewhere \cite{tobepub}.

\begin{figure}
$\begin{array}{c@{\hspace{1cm}}c}
\multicolumn{1}{l} {\mbox{\bf }} &
        \multicolumn{1}{l}{\mbox{\bf }} \\ [-0.53cm]
\epsfxsize=2.1in
\epsffile{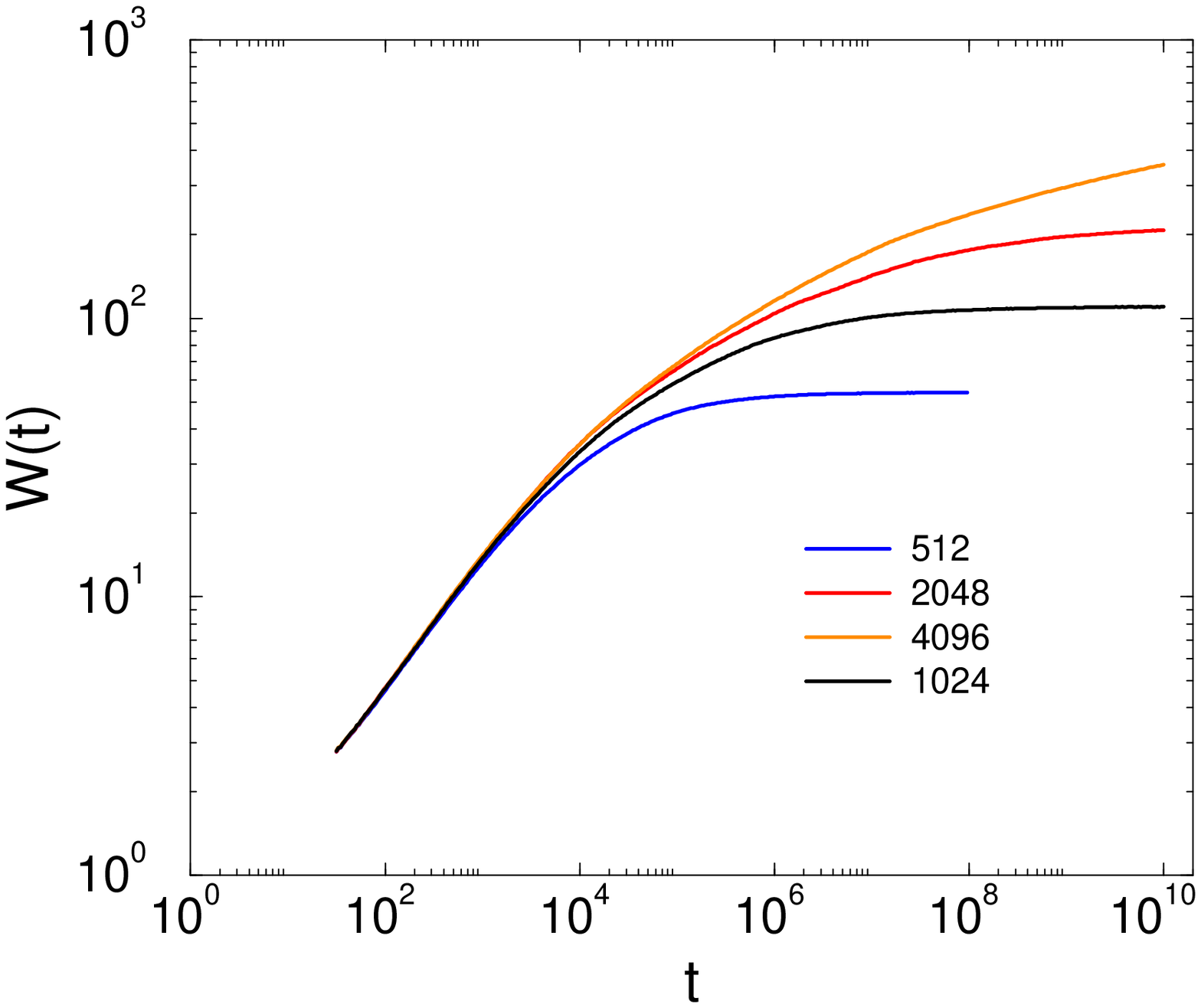} &
\epsfxsize=2.1in
\epsffile{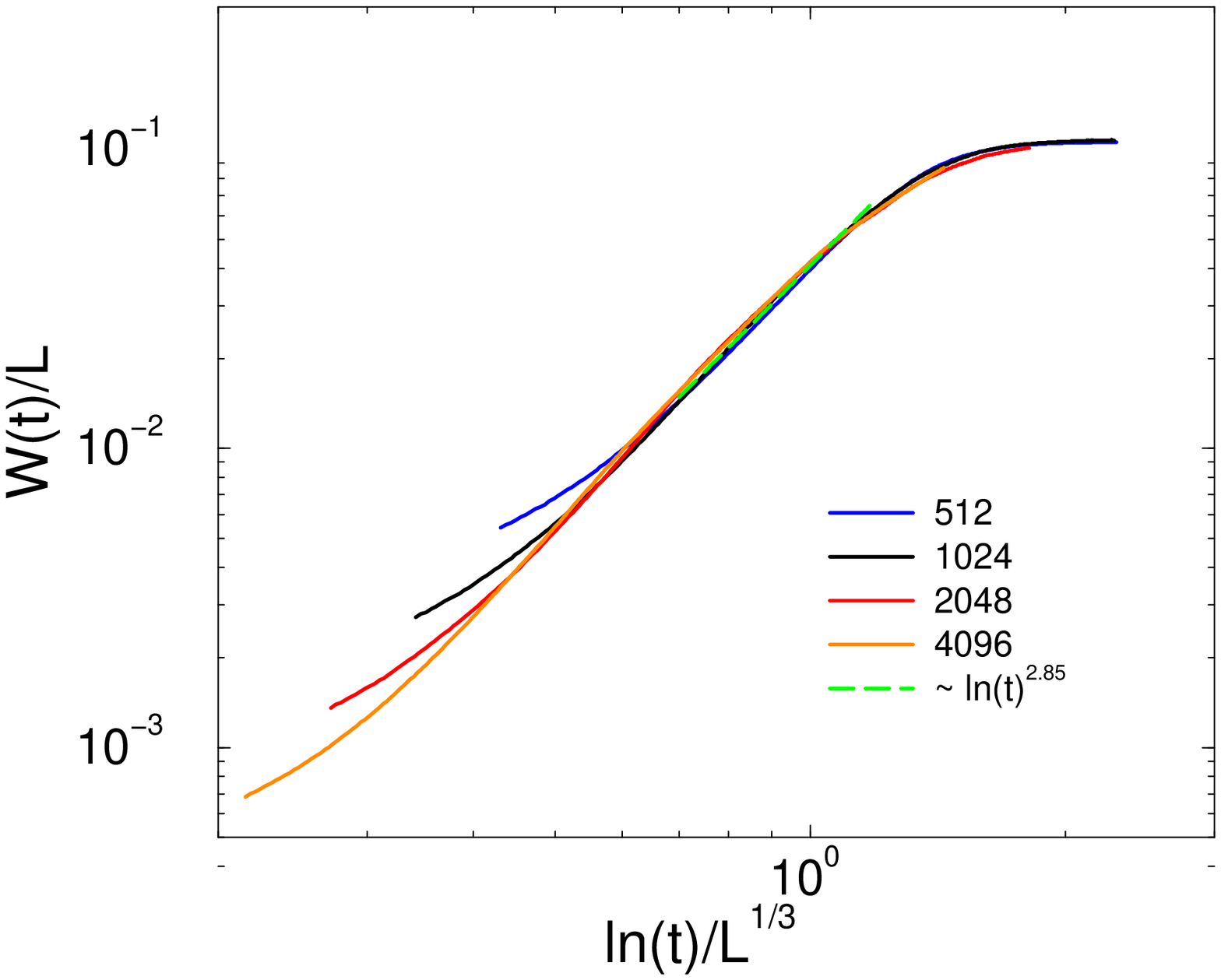} \\ [0.5cm]
\mbox{\bf (a)} & \mbox{\bf (b) }
\end{array}$
\caption{\label{q-ssep}
(a) Surface width of (Q-SSEP) by CUDA simulations for sizes
$L=512, 1024, 2048, 4096$ (bottom to top).
(b) The same, rescaled with the exponents $\alpha=1$, $\psi=1/3$.
The dashed line shows a fitting to $L=4096$.}
\end{figure}

To realize SSEP with disorder (Q-SSEP) it is important to have both
forward and backward jumps of '$1$'-s, otherwise the left-right symmetry
is broken and we find normal scaling instead of an 'activated' one.
As Fig.~\ref{q-ssep}(a) shows an extremely slow convergence
to saturation emerges for $r=0.2$. The evolution of $W(t,L)$ cannot
be described by (\ref{FV-b}), but it follows the activated scaling law
\begin{equation}
W(t,L) \propto \ln(t)^{\tilde\beta}
\end{equation}
as shown on Fig.~\ref{q-ssep}(b). This behavior is expected, 
when the so-called strong disorder fixed point dominates 
in the renormalization group sense \cite{JSI05}.
In this case the dynamics slows down and the relaxation
time is characterized by 
\begin{equation}
\ln(\tau)\propto \xi^{\psi} \ .
\end{equation}
Finite size scaling results in a good data collapse with the 
exponents $\alpha=1$, $\psi=1/3$ and $\tilde\beta=2.85(30)$.
Further results of the data analysis will be discussed elsewhere
\cite{tobepub}.

\subsection{Multi-GPU application by MPI and OpenCL}

Our intention was to find out whether different GPU architectures have 
significant impact on the run-time of simulations and to create a 
multi-card program. To achieve this we favored OpenCL over other 
vendor-dependent APIs. OpenCL is a high-level language, compared to 
NVIDIA's CUDA or AMD's CAL, which are optimal when one develops a 
program for a specific GPU flavor. However creating two implementations 
of the same algorithm, tuning both of them with vendor specific tricks 
is very time consuming, therefore we have chosen OpenCL and wrote 
an algorithm that relies solely on the general aspects of GPU programming
within the framework of programs that execute across heterogeneous 
platforms consisting of CPUs, GPUs, and other processors \cite{khronos} 
and let the compilers to do the optimizations.
        
\begin{figure}[ht]
\begin{center}
\epsfxsize=120mm
\epsffile{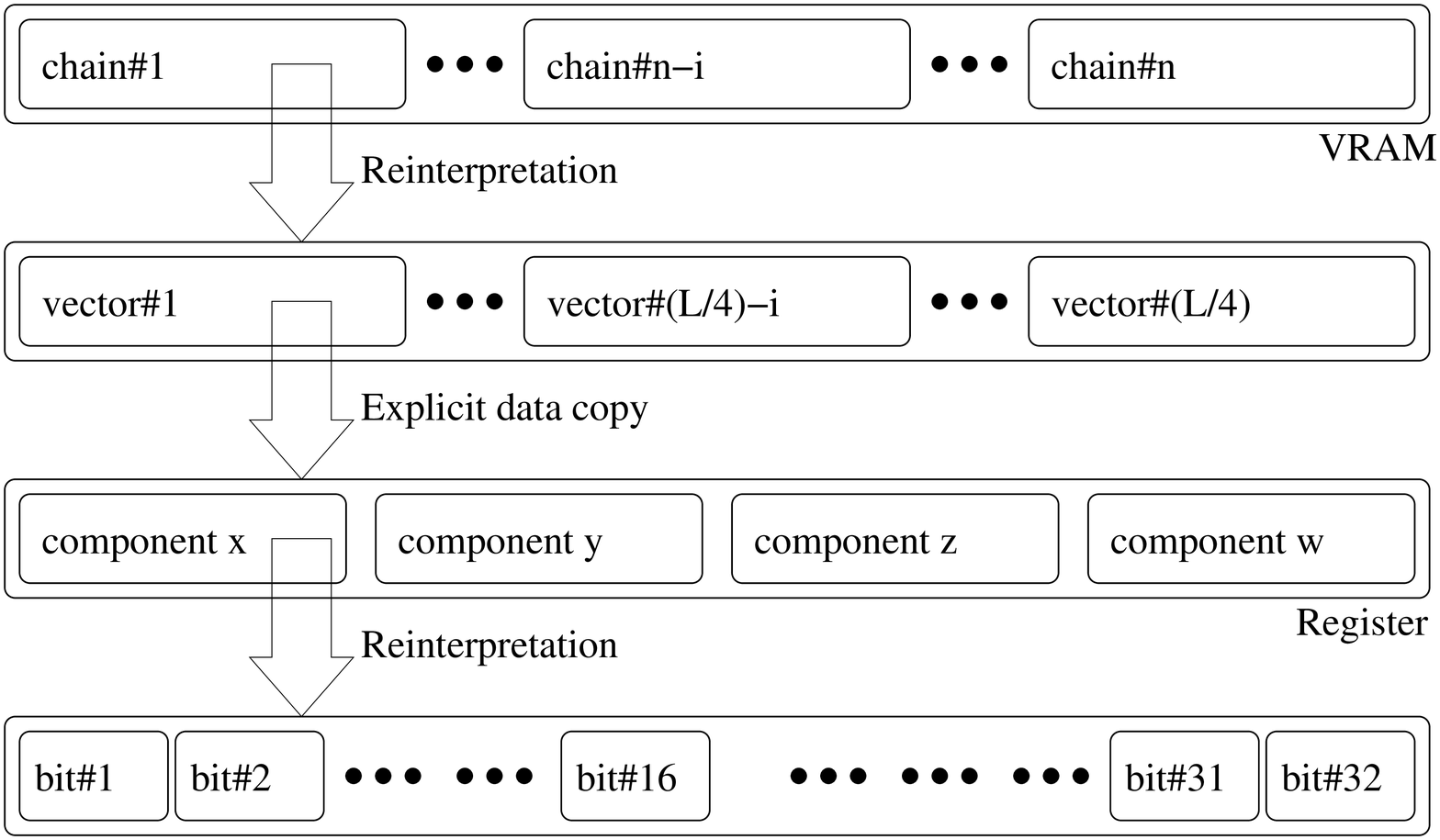}
\caption{Memory map of data in the OpenCL algorithm.} \label{clmap}
\end{center}
\end{figure}

The algorithm used in this program is similar to that of the
previously explained CUDA code, however the implementation differs 
at many points. It is similar, since it uses bit level coding 
of the chain, however the state of lattices sites and the reduced 
probability locations are stored separately. They are not inside the 
same array of characters, but each of them is stored in an 
array of its own (see Fig.~\ref{clmap}). 
This saves space from the shared memory in case of ASEP, allowing 
one to simulate longer chains, if needed. Furthermore, it has a second 
collateral benefit, which ultimately abolishes the need to use 
shared memory, lifting the limit imposed by the size of the shared 
memory of a Compute Unit. This provides a limitation on $L$ only by the
total VRAM size accessible to a GPU.

As we saw earlier, updating lattice sites is completely independent at 
odd and even time-steps, hence we can organize our data, that allows 
processing the chain by vector operations. This optimization 
was used to utilize AMD GPU capabilities, namely the advantage coming 
from the fact that Stream Cores located in AMD GPUs are 4 unit wide 
vector-processors (with a fifth Special Function Unit, inaccessible
directly). This will not hinder NVIDIA cards, they have to deal with 
vector and matrix operations, unrolled to scalar ones either by the 
driver (in case of DirectX applications), 
or by the compiler (in the case of OpenCL applications).

The second change in the implementation is related to another optimization. 
It is a general GPGPU practice, that one tries to avoid flow control 
whenever possible. When there is no actual difference of the program flow 
in two branches of a control statement, besides the value of a single 
variable predication should be used. 
One should keep in mind that predication does not save computation required 
for the evaluation. It only removes the need of branching, which program 
flow to enter, hence it decreases the idle ALU time. 
This might seems to be an over-optimization, but practice shows that 
flow control throws back the performance and should be avoided whenever
possible.

Although less important in long simulations, if the sampling is done
at rare time steps (\ref{sampling}), the output $W^2(t)$ is 
computed in a parallel manner on the GPU device. It is possible to 
add second moments of the height distribution with the proper correction, 
explained in \cite{algo}.
The method implemented is a mixture of the two-pass algorithm and 
the pairwise method. This is useful if multiple, long chains are 
calculated, when it takes longer to collect data.

As it has been stated before the limitation imposed by the size of the 
shared memory can be lifted if we avoid its usage. This is both a
benefit and a drawback of high-level implementations, like OpenCL, 
that the explicit program behavior is masked from the programmer. 
Early implementations showed, that the program ran well, when more 
shared memory was allocated, than what was available to the card. 
This is due to the fact that even if excess data are stored in the VRAM, 
it is still within the shared memory name-space and the compiler does 
not notify the programmer about the excessive memory usage. 
Benchmarks with this implementation showed that run-time did not 
increase, when some of the data was read from device memory.  

\begin{figure}[ht]
$\begin{array}{c@{\hspace{1cm}}c}
\multicolumn{1}{l} {\mbox{\bf }} &
        \multicolumn{1}{l}{\mbox{\bf }} \\ [-0.53cm]
\epsfxsize=2.1in
\epsffile{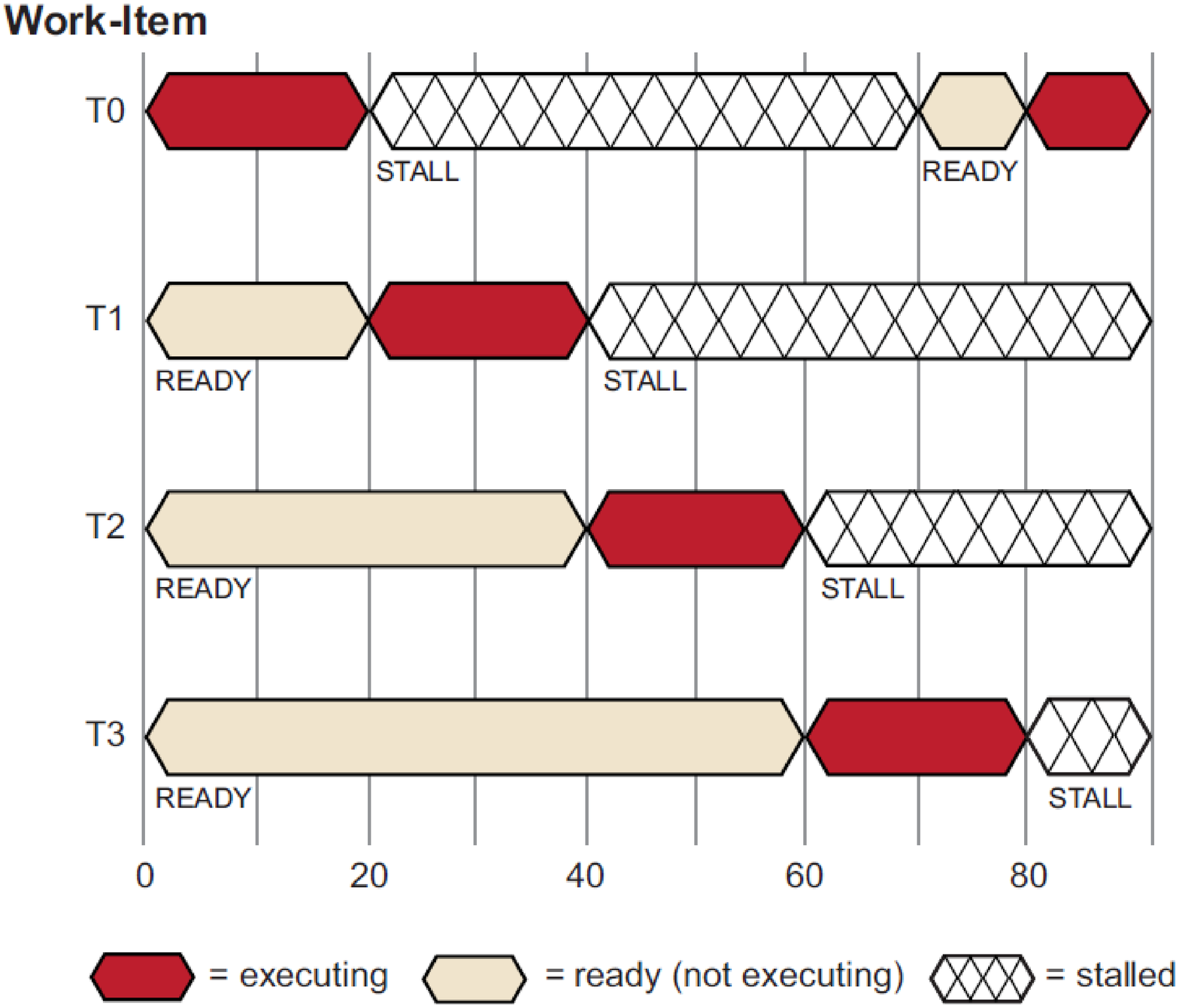} &
\epsfxsize=2.1in
\epsffile{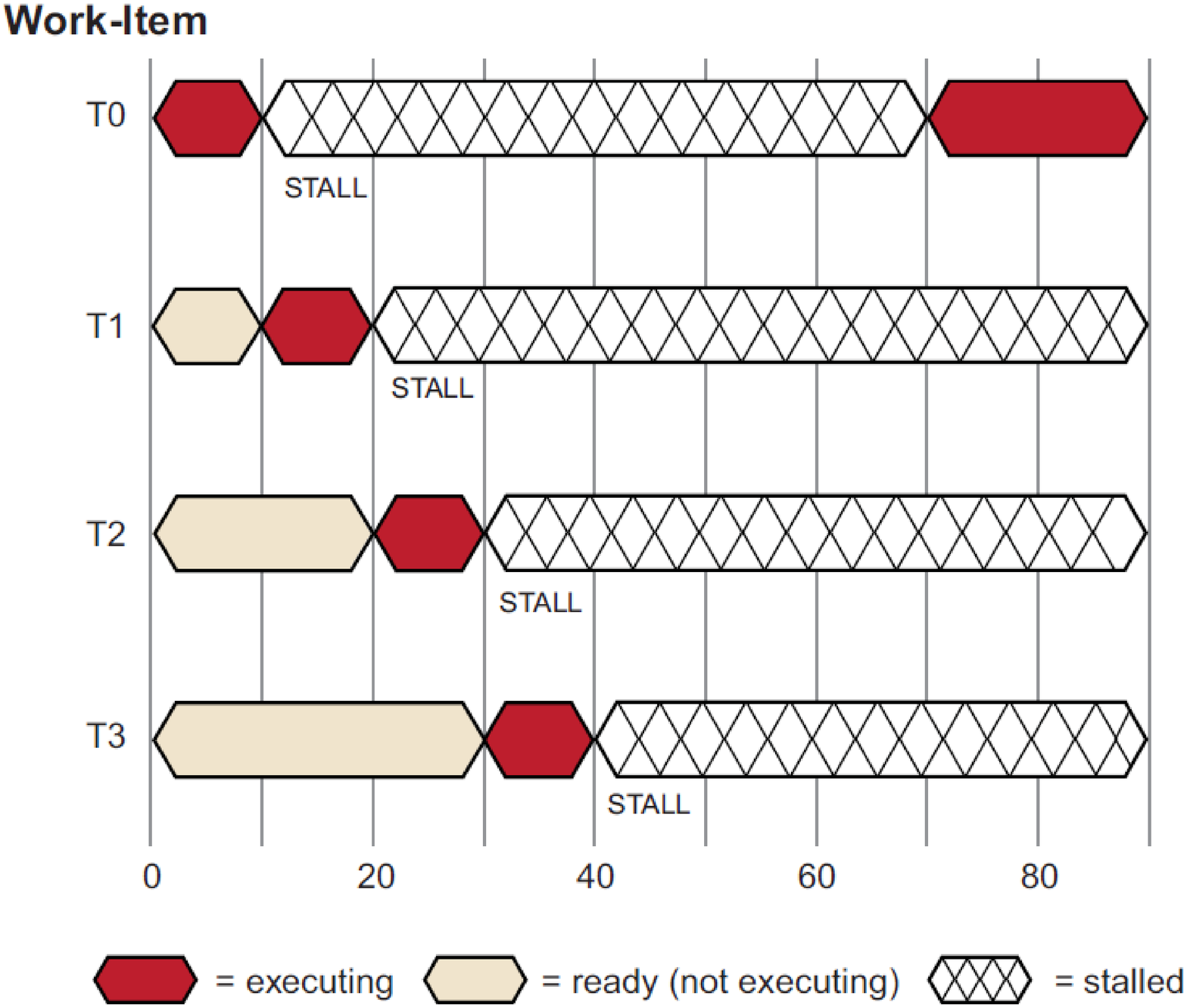} \\ [0.5cm]
\mbox{\bf (a)} & \mbox{\bf (b) }
\end{array}$
\caption{Latency hiding among threads (T0,T1,T2,T3) for
vectorized (a) and non-vectorized execution\label{latency}}
\end{figure}

The two diagrams on Fig.~\ref{latency} help us to understand 
why reorganizing data removes the advantage of shared memory usage. 
The diagrams show how a PE hides read latencies by thread 
scheduling inside a work-group. Dark red parts show useful work, 
beige parts show idle threads waiting for execution and 
crossed parts show idle thread time due to a memory read. 
When a thread requires data outside of its registers, 
the execution stalls until data arrives. If a stall occurs, 
the execution is given to another thread in the work-group, 
until it reaches a read command. If data are fetched in too 
small portions compared to the amount of time required to 
process it, scheduling won't be able to hide read latencies, 
the useful (red) work time will be too short to make up for 
the time needed to read.
If lattice states are stored in vectors of integers, 
one vector read operation requires $128$ lattice points 
to be updated (see Fig.~\ref{latency} (a)). In comparison,
if every lattice point is stored in a separate character, 
a read command is imposed in between lattice updates 
(see Fig.~\ref{latency}(b)).

The main reason for the OpenCL implementation was to investigate if
this kind of problem fits clusters of GPUs. Solving the problem on a 
cluster can be approached in multiple ways. Using multiple GPUs one can 
simply increase the number of chains simulated in parallel to reduce
statistical errors.  Multi-cards can also be used to simulate larger 
chains, spanning over more than one GPUs. 

Implementing the latter idea would be an effort in vain for
the following reasons. The algorithm should copy (a few bits) at 
every lattice sweeps.
In order to optimize the work-time/data-fetch ratio by the vectorization, 
the ratio of inter-lattice communication and lattice sweep time is fixed. 
Even inside shader codes, the most expensive operations are synchronization 
(sync) commands. One sync command takes roughly 2 magnitudes higher 
execution time, than regular ADD and MUL commands. 
Sync commands outside a GPU, but inside the computer are yet again 
roughly 2 magnitudes more costly, while synchronizing over a network 
takes even more time. When one sweep of a lattice points can be
done very quickly, the efficiency of the implemented algorithm 
would depend on the amount of sync and copy commands. 

For this reason the OpenCL implementation avoids shared memory usage
and one can consider large chains (typically $L \sim O(10^6)$).
The multiple card processing capacity is exploited just to increase 
statistics, which is very useful, because a large number of independent 
disorder realization is necessary in these problems.
On the other hand in higher dimensions we will need more memory 
than what is accessible by a single GPU, so the other approach must be
followed. 

OpenCL does not have such a long history as CUDA does, 
compilers are still evolving and not all functionality of the 
standard are supported by them. This leads to slight modifications 
in the code to make it cross-vendor ready. Furthermore, different 
architectural capabilities require to create an implementation, which 
uses the 'greatest common denominator' of capabilities, so to say.

Note, that the CUDA and OpenCL implementations use different random 
number generators. The CUDA code uses gpu-rng, while the OpenCL utilizes a
Mersenne-Twister type of generator, slightly modified to fit to the 
problem. The difference in the performances due to these generators 
were not inspected.

We compared the codes by simulating TASEP chains without disorder.
The performance of the OpenCL implementation on NVIDIA cards is
inferior to that of the CUDA implementation, (at least for $L < 8K$)
due to reasons encountered previously. However, higher dimensional 
simulations will most likely have to use the OpenCL like data
structure, due to the very limited size of the shared memory.

\begin{figure}[ht]
$\begin{array}{c@{\hspace{1cm}}c}
\multicolumn{1}{l} {\mbox{\bf }} &
        \multicolumn{1}{l}{\mbox{\bf }} \\ [-0.53cm]
\epsfxsize=2.1in
\epsffile{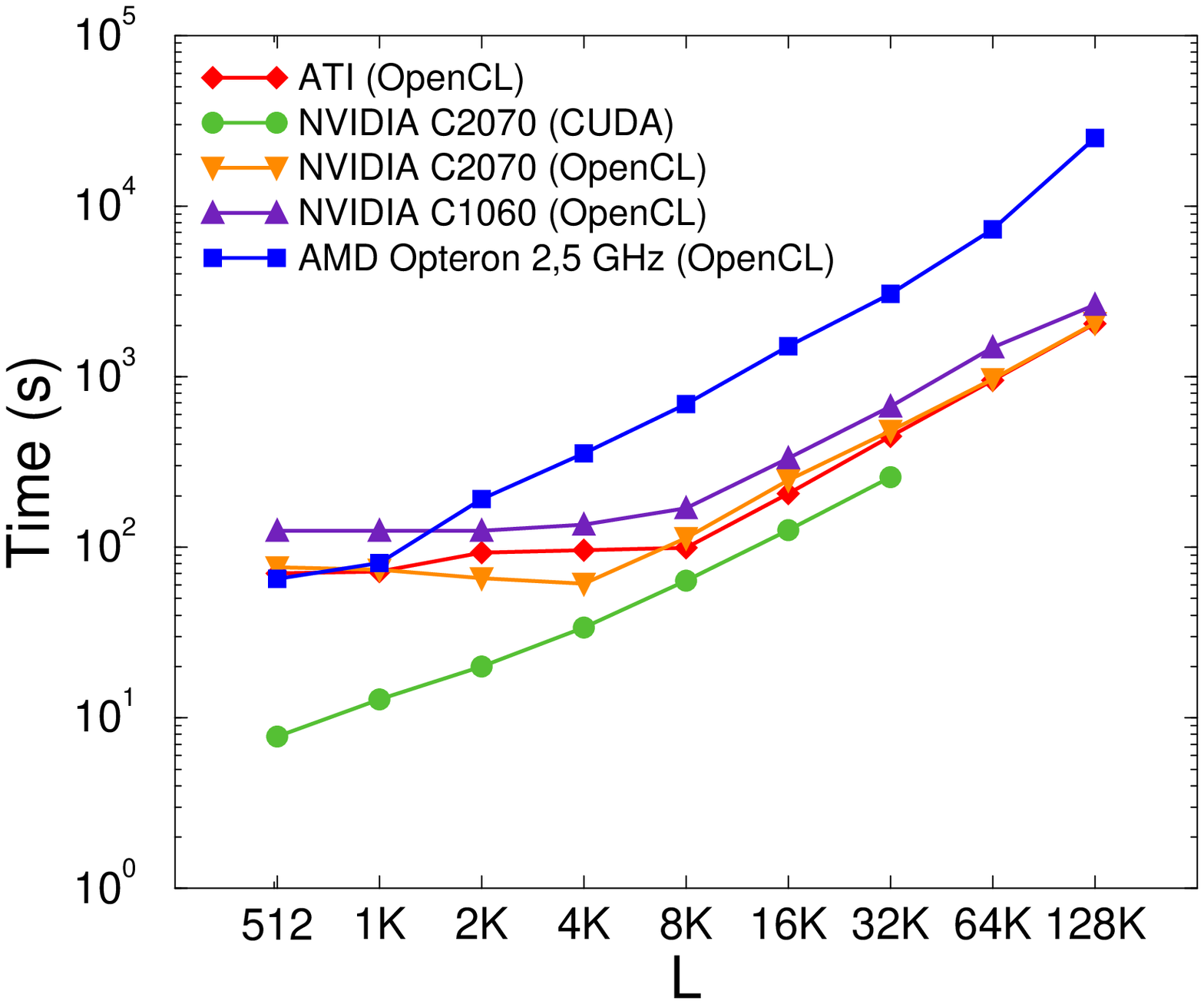} &
\epsfxsize=2.1in
\epsffile{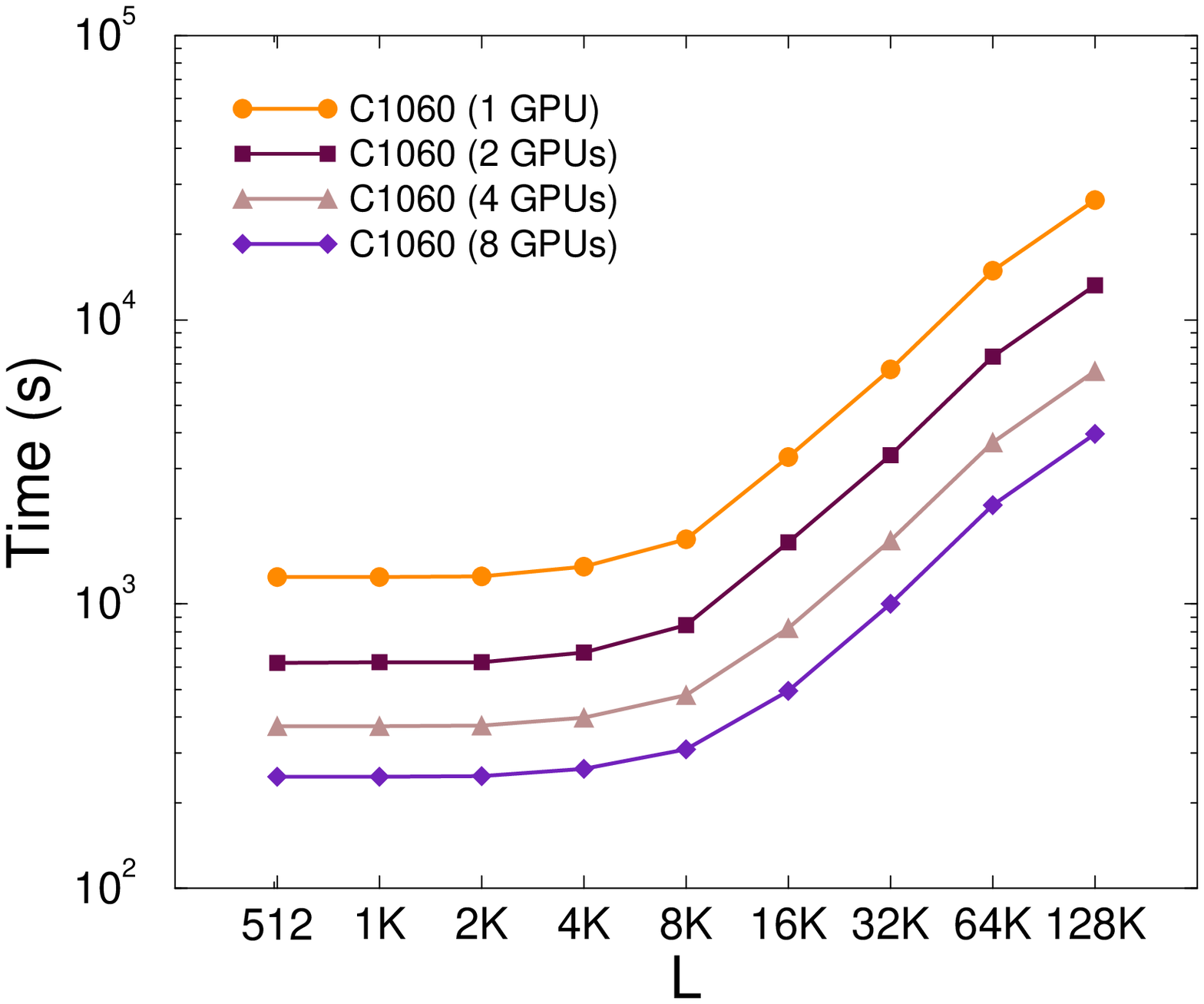} \\ [0.5cm]
\mbox{\bf (a)} & \mbox{\bf (b) }
\end{array}$
\caption{\label{per} Run-time comparisons for different implementations 
(CUDA, OpenCL) of TASEP on different platforms (single ATI GPU, 
single NVIDIA GPU, single cluster node containing 16 CPU cores) (a) 
and for the OpenCL implementation using 1, 2, 4 and 8 NVIDIA C1060
(b).}
\end{figure}	

The performance scaling of the algorithm is nearly optimal, 
as one can expect it, taking into account that different 
threads run independently from each other. 
Intimacies between the MPI and GPU drivers cause that ideal 
scaling cannot be achieved. 
The OpenCL implementation was tested on NVIDIA C1060 and C2070 cards, 
on ATI Hemlock cards and also on conventional cluster nodes 
with 4 AMD Opteron quad-core CPUs. 
The CUDA implementation was tested on the NVIDIA cards. 

The common task to be done was running the simulations up 
to $t_{max}=10^5$ MCs for $560$ realizations. Although the
the architecture of Tesla C1060 GPU is considerably different 
from that of TESLA C2070 we could use the same algorithms.
OpenCL is responsible for rearranging memory stores/loads and 
ALU instructions to fit the architecture best. 
Differences between C1060/C2070, such as the L2 cache remain 
unexposed to the user, are not accessible directly. The
compiler and the HW decide how to use them. In case of CUDA we used
different thread numbers for C1060/C2070 to achieve the best 
performance for a each. The algorithms could be optimized further,
if the development time was unlimited, but our code is far from 
being a first-approach, exploits both basic and nontrivial 
aspects of GPU programming \cite{khronos,nvidia}.

As one can see on Fig.~\ref{per}(a) the run-time ($T$(seconds)) 
initially does not change in the case of OpenCL ran on GPUs 
as $L$ is increased. 
This is because GPUs are not fully utilized on short chains
and CUDA cores are idle due to the lack of work. 
Run-times begin to increase as all PE-s start to work. 
The constant time region is longer for the Fermi based Teslas, 
due to the higher number of CUDA cores inside a Compute Unit. 
The OpenCL version, running on 16 CPU cores does not have the 
constant part for smaller chain lengths, but the scaling is 
quite similar to that of the GPUs. 

The CUDA implementation on the Fermi cards runs significantly faster, 
especially for smaller chain lengths. The run-times grow almost 
linearly ($T \propto L^{0.9}$) for all sizes: $512 \le L \le 32K$, 
that can be simulated due to the shared memory limitation.
\begin{figure}[ht]
$\begin{array}{c@{\hspace{1cm}}c}
\multicolumn{1}{l} {\mbox{\bf }} &
        \multicolumn{1}{l}{\mbox{\bf }} \\ [-0.53cm]
\epsfxsize=2.1in
\epsffile{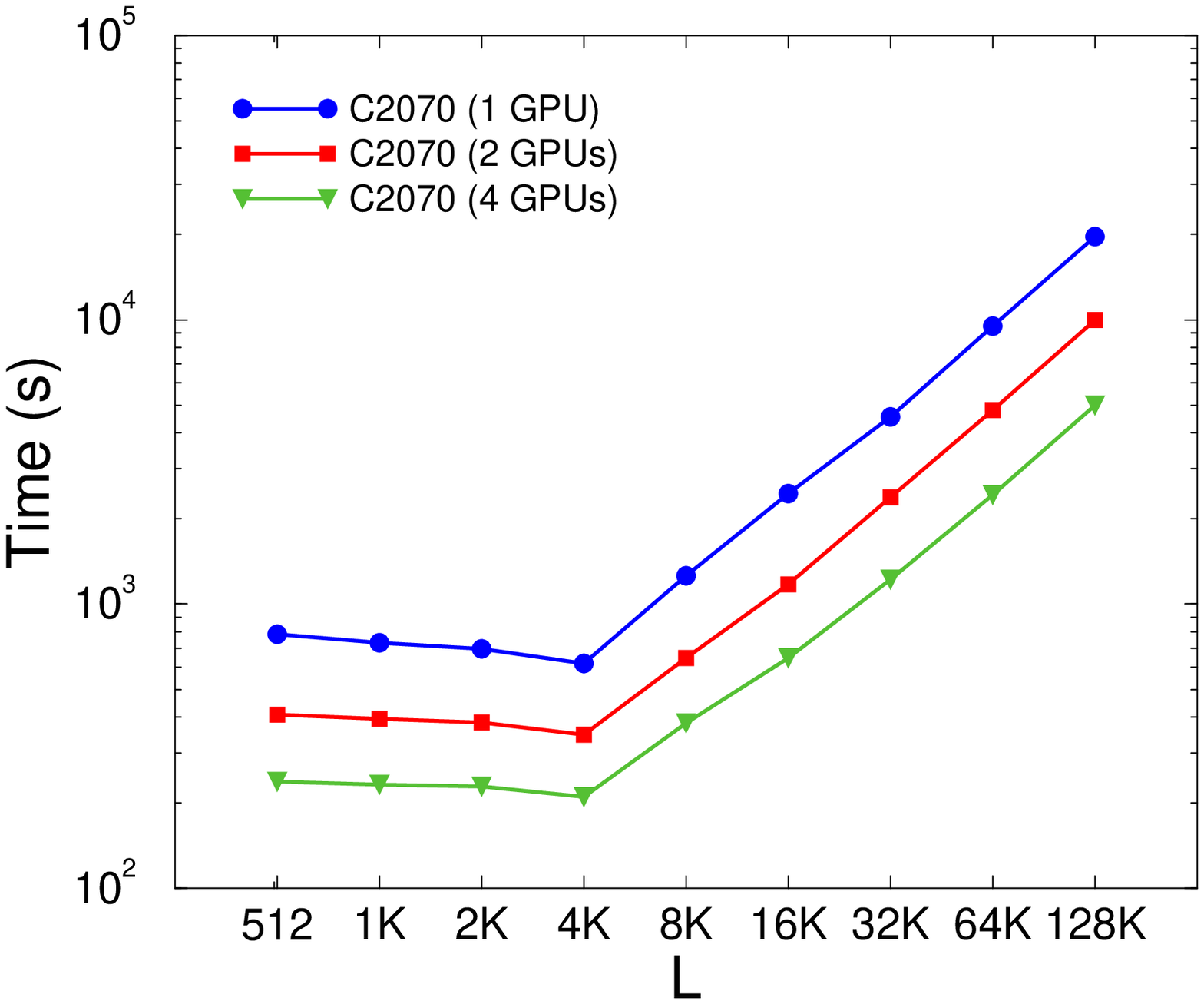} &
\epsfxsize=2.1in
\epsffile{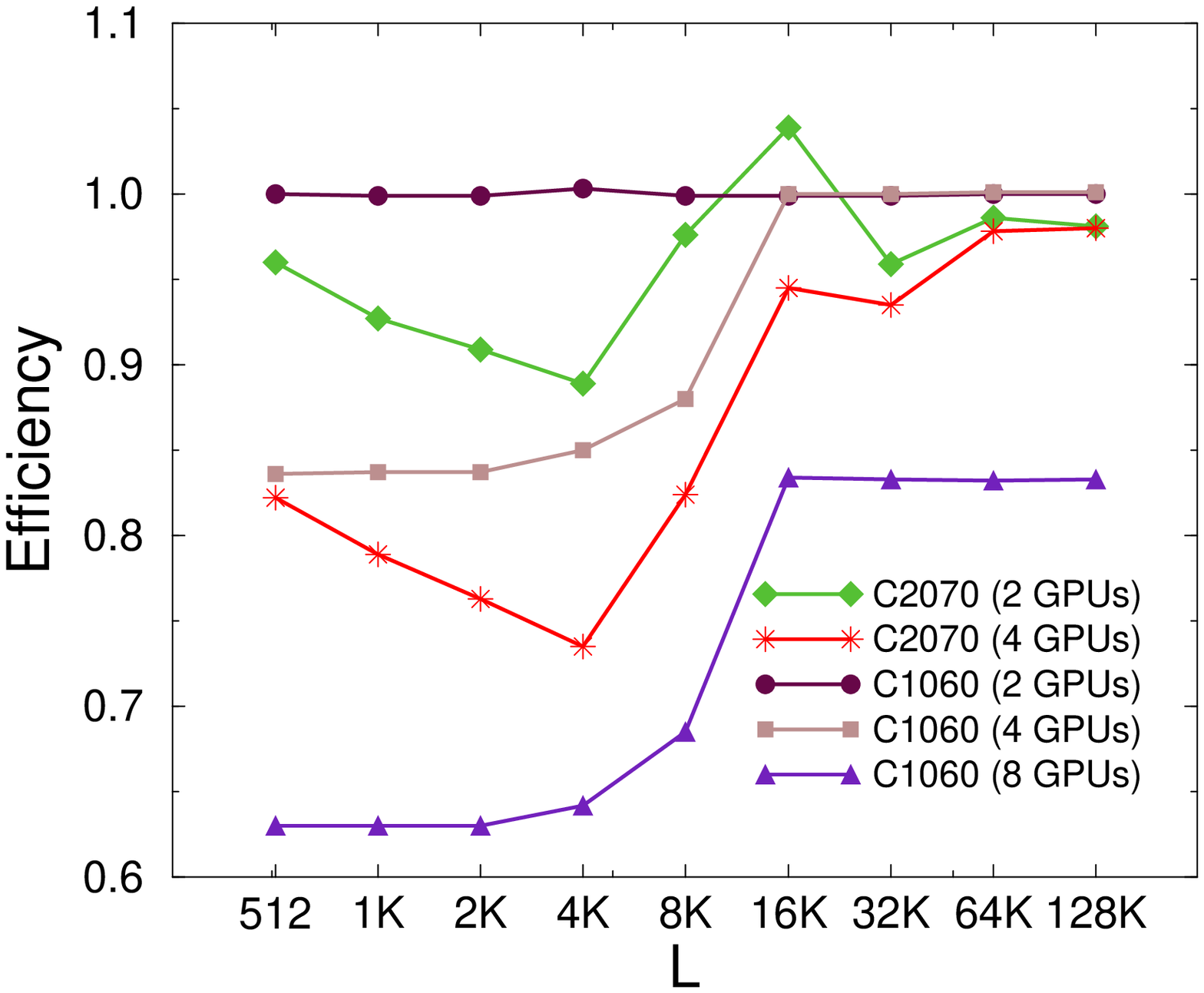} \\ [0.5cm]
\mbox{\bf (a)} & \mbox{\bf (b) }
\end{array}$
\caption{\label{per2} (a) Run-time comparisons for the OpenCL 
implementation using 1, 2 and 4 NVIDIA C2020.
(b) Efficiency on 2, 4 and 8 NVIDIA C1060 and 2 and 4 NVIDIA C2070.
The peak for $L =16K$ on NVIDIA C2070 comes from a small delay on 
the server in the referenced 1-GPU simulation.}
\end{figure}	

Multi-GPU run-time comparison was done on a limited number of 
NVIDIA GPUs, since only 4 C2070 and 8 C1060 NVIDIA cards have been 
available for us. 
Fig.~\ref{per}(b) shows the scaling on the 8 Teslas (C1060); 
while Fig.~\ref{per2}(a) displays the same for the 4 Fermi
cards (C2070). As one can see these are nearly optimal, 
the curves go parallel for all tested chain lengths and the speedup
grows almost linearly with the number of cards. This is far from trivial
in case of parallel algorithms running on multiple cards.
The OpenCL frame will be very useful for further applications of this type.
In Fig.~\ref{per2}(b) one can see that the efficiency, defined as the 
speedup per node, of the multi-GPU implementations depends 
on the number of cards, as well as on $L$ in a non-monotonic way.
This enables one to optimize simulations with limited resources.

\section{Two-lane ASEP model simulations}

Next we turn to the numerical study of a two-lane model, with opposite, 
unidirectional motion in the lanes and with quenched, random inter-lane 
crossing probabilities \cite{J10}.
This arrangement can be thought of a simplified model of cars or motors 
moving along two oppositely oriented roads or filament tracks.
The motion of a single particle in this environment has been found to
exhibit enhanced diffusion coefficient compared to that of the 
symmetric random walk \cite{KL05}. 
This type of active diffusion can be realized experimentally 
and also present {\it in vivo} \cite{G04}.
Theoretical studies have concluded that, the steady-state behavior 
is richer than that of the one-lane ASEP, furthermore interesting 
phenomena take place in such systems, like synchronization or 
localization of density shocks.

\begin{figure}[ht]
\begin{center}
\epsfxsize=70mm
\epsffile{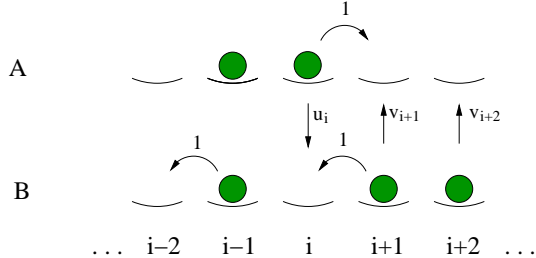}
\caption{The bidirectional, two-lane model defined in 
ref~\cite{J10}} \label{bifig} \end{center}
\end{figure}

The bidirectional, two-lane model is built up from two TASEP rings,
in which particles move forward in both lanes deterministically,
to the right in $A$ and to the left in $B$ (see Fig.~\ref{bifig}),
and random lane-crossing probabilities: $u_i$, $v_i$.
These are quenched, random variables with bimodal distribution 
(\ref{bimodal}).
We have been interested in the dynamical behavior of this model by
determining the exponent $z$ via finite size scaling. 
In the steady-state the total current of particles 
($c(t\to\infty, L$) vanishes due to the left/right symmetry 
and its fluctuation $<c(t,L)^2>$ is expected to grow linearly 
in time \cite{KL05}.

\begin{figure}[ht]
\begin{center}
\epsfxsize=70mm
\epsffile{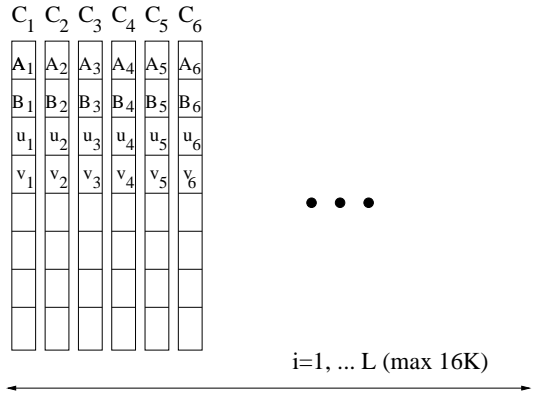}
\caption{Representation of the 2-lane ASEP model by a
vector of characters $C_i$ in the shared memory.} \label{bitek}
\end{center}
\end{figure}

First we realized and tested the SCA version of this model on a CPU, 
second we implemented it by a CUDA program. In order to convert the 
originally random sequential updates to SCA, as in case of the ASEP,
we decreased the in-lane hopping probabilities to $p=0.5$.
One sweep (MCs) of the system consists of the parts executed 
in the following sequence:
\begin{enumerate}
\item Even sub-lattice updates of the lanes $A$ and $B$, 
with probabilities $p=0.5$, 
\item $A\to B$ lane-changes, with probabilities $u_i$, 
\item Odd sub-lattice updates of the lanes $A$ and $B$, 
with probabilities $p=0.5$, 
\item Closing boundaries of $A$ and $B$, with probabilities $p=0.5$, 
\item $B\to A$ lane-changes, with probabilities $v_i$.
\end{enumerate}
All local data are packed in the shared memories, in the vector of 
characters $C_i$ (see Fig.~\ref{bitek}). 
The first bits of this vector contain lane-$A$ states, 
the second bits lane-$B$,  the third and fourth bits ($u_i,v_i$) 
mark the sites, where the crossing probability is reduced from 
$0.8$ to $0.2$. 
These are filled at the at the beginning of time loops by
'$0$'-s and '$1$'-s randomly, with probability $1/2$.
The updates are done by bit-masking and using standard, bit-wise 
(AND,XOR,...) operations in parallel and independently 
by the threads.

We increased/decreased a variable ($J\pm 1$), measuring the current,
by each hopping event along the lanes $A$ and $B$ respectively.
The simulations were started from half filled, random initial conditions 
of system with linear sizes: $L=2^9$, $2^{10}$, $2^{11}$, $2^{12}$. 
Sampling of the current density: $c=(J/L)^2$ was done in the same
way as in case of ASEP simulations (\ref{sampling}). At the end of
a time loop we calculated the total current density fluctuations of the 
system $<c(t,L)^2>$ (the mean value is zero due to the left/right symmetry). 
Averaging was done for $\sim 1000$ samples for each size.

\begin{figure}
$\begin{array}{c@{\hspace{1cm}}c}
\multicolumn{1}{l} {\mbox{\bf }} &
        \multicolumn{1}{l}{\mbox{\bf }} \\ [-0.53cm]
\epsfxsize=2.1in
\epsffile{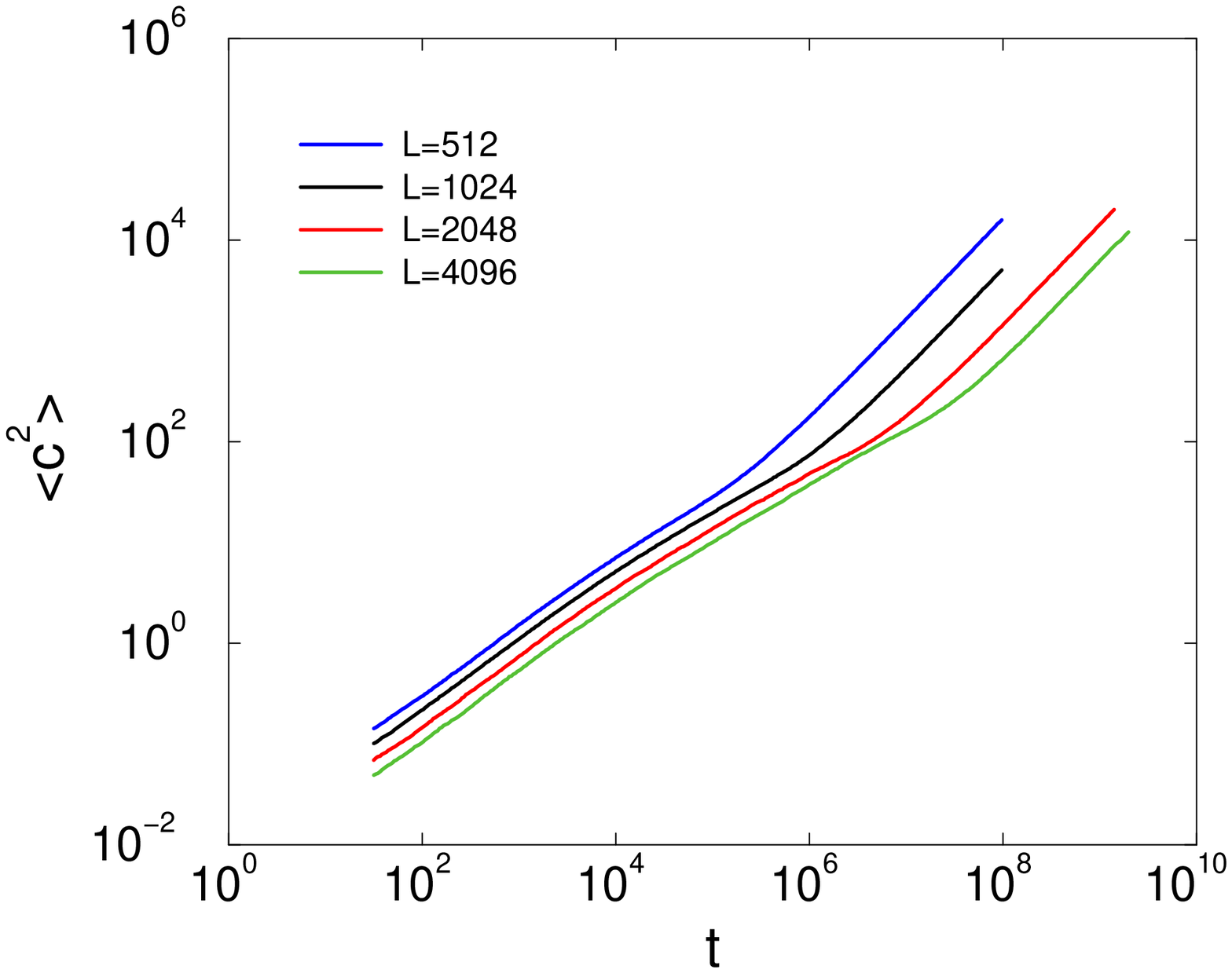} &
        \epsfxsize=2.1in
        \epsffile{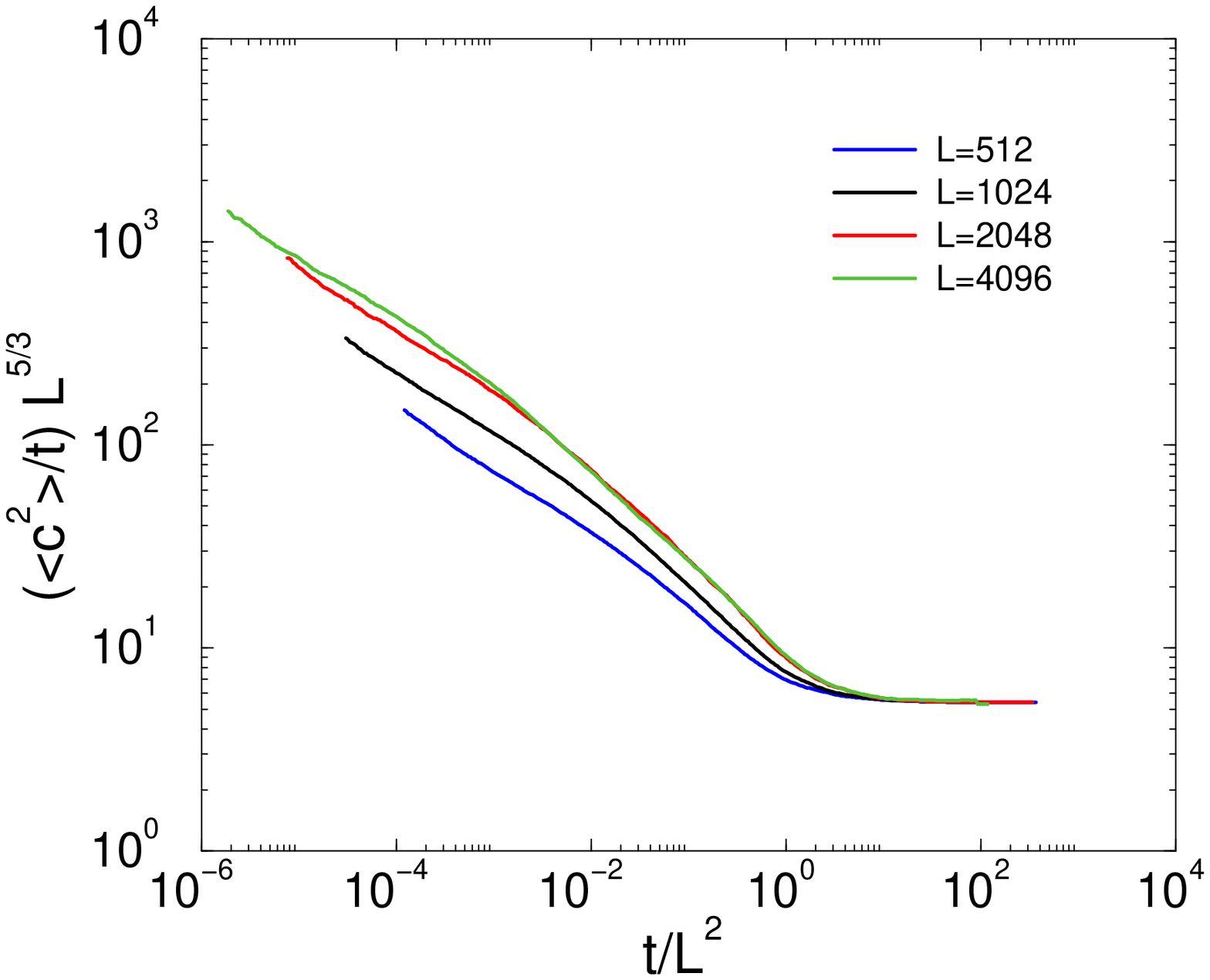} \\ [0.5cm]
\mbox{\bf (a)} & \mbox{\bf (b) }
\end{array}$
\caption{\label{charge}
(a) Fluctuation of the total current in the two-lane model for
sizes $L=512,1024,2048,4096$ (top to bottom).
(b) Finite size scaling and collapse with the same data.}
\end{figure}

As we can see on Fig.~\ref{charge} (a), following an initial transient 
time the shape of each curve changes. The dynamical behavior crosses over 
to a different regime, the steady state, where we can find linear growth 
due to the diffusion like behavior of particles.
Indeed if we divide $<c(t,L)^2>$ by the time, a level-off to constant
value can be observed. To determine the dynamical exponents we performed 
finite size scaling (see Fig.~\ref{charge} (b)).
The best collapse could be achieved by assuming the dynamical exponent
$z=2$, but strong corrections can also be observed. 
A more detailed analysis will be provided in \cite{tobepub}. 

The speed-up of the code on the FX5800 card compared to the reference CPU 
was about $\times 30$. If we allow particles to travel among 
multiprocessor blocks (via the VRAM) we can simulate much larger
sizes, i. e. one sample/GPU. In this case we could reach a 
speed-up $\times 20$.

\section{Conclusions}

We have realized efficient $1d$ lattice-gas simulation algorithms for GPUs
for the first time. This allows us to investigate $1+1$d surface growth or
transport phenomena with extremely long relaxation times. This is typical 
in case of system with quenched random reaction rates. Furthermore in these
system averaging must be carried out over a large number of disorder 
realizations, thus this problem fits the advantages of parallel processors. 
We have confirmed that the SCA version of ASEP is equivalent to
the random sequential one, hence it can be well adapted for simulations
with massively parallel GPU cores. We used two-sub-lattice updates, which
can be generalized to checker-board lattice decomposition in two dimensions.
Our preliminary results for ASEP with quenched disorder are in agreement
with some previous expectations, detailed discussion is in preparation
\cite{tobepub}. 

By coupling two ASEP chains with lane crossing probabilities
we have made the first step towards two dimensional simulations
\cite{asep2dcikk,fzd2010},
as well as we found interesting results for a traffic problem.
We used bit coding to allow efficient memory management, which 
is critical if the fast shared memory of GPUs to be used.

Two similar code realizations have been compered in detail on 
different GPU hardwares.
The CUDA code is faster for smaller system sizes, than the OpenCL one,
but has limitations due to the size of the shared memory and the architecture.
In particular for the simple ASEP problem we could achieve a speedup of 
$\sim100$ as compared to a traditional, present day single CPU. 
This is near the physical limits of the GPU card. 
The multi-GPU (MPI) OpenCL code has been proved to be portable not only
among GPUs of different vendors, but runs on clusters of CPUs as well. 

We have provided detailed scaling analysis of these codes and pointed out
implementation specific questions.
These results warrant for further interesting applications of GPUs
in statistical physics and materials science 
\cite{asepddcikk,patscalcikk,BGS07}.

\section{Acknowledgments}

We thank R. Juh\'asz, I. Borsos, K.-H. Heinig, J. Kelling and N. Schmeisser 
for the useful discussions.
Support from the Hungarian research fund OTKA (Grant No. T77629),
and the bilateral German-Hungarian exchange program DAAD-M\"OB 
(Grant Nos. 50450744, P-M\"OB/854) is acknowledged.
The authors thank NVIDIA for supporting the project with high-performance
graphics cards within the framework of Professor Partnership.

\end{document}